\documentclass[11pt]{article}
\usepackage{aas2pp4,psfig}

\psrotatefirst
\newcommand{\etal}{\hbox{et al.}}
\newcommand{\pmax}{p_{\rm max}}
\newcommand{\lmax}{\lambda_{\rm max}}
\newcommand{\mic}{~$\mu$m}
\renewcommand{\deg}{^{\circ}}

\begin{document}

\title{INTERSTELLAR POLARIZATION IN THE TAURUS DARK CLOUDS: 
       WAVELENGTH DEPENDENT POSITION ANGLES AND CLOUD STRUCTURE NEAR TMC--1}
\bigskip
\author{D. W. Messinger, D. C. B. Whittet and W. G. Roberge}
\affil{Department of Physics, Applied Physics, and Astronomy, 
Rensselaer Polytechnic Institute, Troy, NY 12180}
\centerline{To appear in the Astrophysical Journal, Sept. 20, 1997,
vol 487}

\begin{abstract}
Systematic variations with wavelength in the position angle of interstellar 
linear polarization of starlight may be indicative of multiple cloud 
structure along the line of sight. We use polarimetric observations of two 
stars (HD~29647, HD~283809) in the general direction of TMC--1 in the Taurus 
Dark Cloud to investigate grain properties and cloud structure in this 
region. We show the data to be consistent with a simple two-component model, 
in which general interstellar polarization in the Taurus Cloud is produced 
by a widely distributed cloud component with relatively uniform magnetic 
field orientation; the light from stars close to TMC--1 suffers additional 
polarization arising in one (or more) subcloud(s) with larger average grain 
size and different magnetic field directions compared with the general trend. 
Towards HD~29647, in particular, we show that the unusually low degree of 
visual polarization relative to extinction is due to depolarization 
associated with the presence of distinct cloud components in the line of 
sight with markedly different magnetic field orientations. Stokes parameter 
calculations allow us to separate out the polarization characteristics of 
the individual components. Results are fit with the Serkowski empirical 
formula to determine the degree and wavelength of maximum polarization. 
Whereas $\lmax$ values in the widely distributed material are similar to the 
average (0.55~\micron) for the diffuse interstellar medium, the subcloud in 
line of sight to HD~283809, the most heavily reddened star in our study, has 
$\lmax \approx 0.73$~\micron, indicating the presence of grains $\sim 30\%$ 
larger than this average. Our model also predicts detectable levels of circular 
polarization toward both HD~29647 and HD~283809.
\end{abstract}

\section{INTRODUCTION}

The dark cloud complex in Taurus provides a particularly suitable 
laboratory in which to study the properties of interstellar dust 
grains in quiescent environments. As it is relatively 
nearby ($\sim 140$\,pc) and $\sim 20^\circ$ from the Galactic plane, 
extinction in this area of sky arises almost exclusively within the 
cloud itself (Elias 1978; Strai\v{z}ys \& Meistas 1980; Kenyon, 
Dobrzycka \& Hartmann 1994). The physical and chemical properties of both 
the dust and gas are reasonably well constrained. The cloud has a 
filamentary structure (e.g., Cernicharo \etal\ 1985), containing 
embedded dense cores detected in line emission of gas phase molecules 
such as NH$_3$ and CS (e.g., Myers \& Benson 1983; Zhou \etal\ 1989). 
The cores have typical masses of a few M$_\odot$, gas temperatures of 
$\sim 10$\,K, and densities $\sim$ $10^4$\,cm$^{-3}$ (e.g., Wilson \& 
Minn 1977), and are active sites of low-mass star formation (e.g., 
Lada \etal\ 1993). A discontinuity between dust properties in the 
outer layers and dense inner regions of the cloud has been 
demonstrated. Grains in the outer layers (at visual extinctions 
$A_V < 3$) have optical properties similar to those of dust in the diffuse 
interstellar medium (Vrba \& Rydgren 1985; Kenyon \etal\ 1994), whereas 
infrared spectroscopy demonstrates the presence of icy mantles on the 
grains in regions of higher extinction (Whittet \etal\ 1988, 1989; 
Chiar \etal\ 1995). 

As in other regions of the interstellar medium (ISM), polarization of 
background starlight in the Taurus dark cloud (TDC) is caused by the 
alignment of spinning grains with respect to the ambient magnetic field 
(see Roberge 1996 for a recent review of the physics of grain alignment). 
The rapid Larmor precession of the grains' angular momenta 
about magnetic field lines results in alignment correlated with the 
direction of the magnetic field (Martin 1971).
Observations of linear polarization may thus be used to 
map the magnetic field in the plane of the sky; results for the
TDC (Moneti \etal\ 1984; Tamura \etal\ 1987; Myers \& Goodman 
1991; Goodman \etal\ 1990, 1992) show that the distribution of polarization 
vectors is fairly uniform over the entire cloud on a macroscopic 
scale. However, Goodman \etal\ (1995) cautioned that maps of optical
polarization may not be indicative of conditions deep within 
molecular clouds if the grains in high-density regions are not efficient 
polarizers. Gerakines, Whittet, \& Lazarian (1995) showed that the
polarization efficiency, $p/A$, decreases rapidly with increasing extinction 
within the TDC. This general trend might be caused by sensitivity 
of the alignment mechanism to physical conditions; however, it could also 
be due, at least in part, to depolarization effects associated with 
small-scale structure within the cloud. The relationship between 
polarization measurements, magnetic field and cloud structure is therefore 
of great importance to our understanding of cloud 
fragmentation, collapse, and star formation.

A good candidate for the study of depolarization effects associated with 
small-scale structure is the line of sight to the reddened B-type star 
HD~29647 (see Table~1 for basic information and references). Fortuitously, 
this star lies behind the outer regions of the TMC--1 condensation within the 
dark cloud (Crutcher 1985). Depolarization in this line of sight is 
suggested by the fact that the degree of linear polarization is much less 
than expected for the degree of extinction, compared with other stars in 
the region (Whittet \etal\ 1992). The polarization measurements also show 
significant rotation of the position angle with wavelength (by $23\deg$ 
between 0.35 and 2.0~\micron; see Figure~1b), a phenomenon most naturally 
explained in terms of the presence of two or more cloud components with 
differing grain size distributions and magnetic field directions (Coyne 
1974). Another star behind TMC--1, HD~283809, shows similar but less 
extreme behavior. 

In this paper, we analyze the wavelength dependence of polarization 
towards HD~29647 and HD~283809 with reference to a `control' star 
(HD~283812) with `normal' polarization properties. First, we describe 
the nature of the region and summarize previous work 
on the optical properties of the dust in the line of sight. The method 
of analysis is described in \S~3. Results are presented and discussed 
in \S~4, and our conclusions are summarized in \S~5.

\section{DESCRIPTION OF THE REGION}

HD~29647 lies at a distance of $170\pm20$\,pc, and therefore appears to be 
situated just behind the TDC region at $140\pm10$\,pc (Crutcher 1985). 
Contributions from foreground and background material are likely to be 
small, the visual extinction of the star ($A_V \approx 3.7$) arising almost 
entirely within the TDC itself. The line of sight to HD~29647 intersects 
Heiles Cloud~2 within the TDC at a point $\sim 8'$ north of the 
northernmost clump of dense molecular gas in TMC--1 (see Figure~1 of 
Crutcher 1985). Schloerb \& Snell (1984) describe Heiles Cloud~2 as a 
rotating ring that is fragmenting as it collapses, and interpret TMC--1 as 
a dense filament resulting from this collapse\footnote{Tamura \etal\ (1987) 
review other explanations for the velocity structure in the line of sight, 
which include the possibility of local disturbances resulting from a shock 
wave or cloud-cloud collision.}. Observations of $^{13}$CO line emission at 
the position of HD~29647 clearly reveal the presence of two distinct 
components, at radial velocities of 5.1 and 6.5\,km/s relative to the LSR 
(see Schloerb \& Snell 1984 and Crutcher 1985; the spectrum in the upper 
left frame of Schloerb \& Snell's Figure~4 corresponds to the position of 
HD~29647). Data for molecules such as CN, HCN and HCO$^+$, which trace 
denser material, are consistent in velocity with the 5.1\,km/s component, 
which is therefore assumed to correspond to clump material. These 
observations strongly support our adoption of a two-component model, 
such as a clump embedded in lower density material, for the dust towards 
HD~29647. HD~283809 lies in the same general region ($\sim 6'$ SE of 
HD~29647) and has substantially higher extinction ($A_V \sim 5.8$), 
suggesting a greater contribution from the clump. Unfortunately, no 
detailed information on molecular line velocities is available specific 
to this line of sight. 

The wavelength dependence of interstellar extinction in the TDC 
has been studied in the visible and near infrared by several authors 
(Strai\v{z}ys, Wisniewski \& Lebofsky 1982; Strai\v{z}ys, \v{C}ernis \& 
Hayes 1985; Vrba \& Rydgren 1985; Kenyon \etal\ 1994; Whittet \etal\ 1997), 
with generally consistent results. Stars with relatively low extinctions 
($A_V<3$) have normal extinction curves, characterized by values of the 
ratio of total to selective extinction ($R_V = A_V/E_{B-V}$) close to the 
average of $3.1$ for diffuse regions of the ISM. HD~29647 and HD~283809 
have values of $R_V$ (Table~1) somewhat above this average, indicating a 
trend towards larger average particle size. This is consistent with the 
presence of denser material in these lines of sight, in which grain growth 
processes become more rapid. The extinction curve of HD~29647 shows more 
dramatic differences in the ultraviolet compared with the average 
for the diffuse ISM (Cardelli \& Savage 1988). The mid-ultraviolet 
extinction bump centered near 2175~\AA\ is substantially broader, weaker, 
and shifted to shorter wavelength, whereas the amplitude of the far 
ultraviolet extinction rise is relatively high. The origin of these 
differences is not well understood, but it seems likely that growth 
processes in dense material along the line of sight have somehow modified 
and/or depleted the carrier of the bump. No information is available on 
the extinction curve of HD~283809 in the satellite ultraviolet. However, 
Strai\v{z}ys \etal\ (1985) find appreciable differences between HD~283809 
and HD~29647 in the blue to near ultraviolet, leading them to conclude that 
the properties of the small-grain population are different in these lines 
of sight despite their proximity in the sky.

The wavelength dependence of interstellar polarization in the TDC 
has been studied by Whittet \etal\ (1992, 1997). Results are well 
characterized by the `Serkowski law' (Serkowski 1973; Serkowski, Mathewson 
\& Ford 1975; Wilking \etal\ 1980, 1982; Whittet \etal\ 1992; see \S~3 below). 
Values of the wavelength of maximum polarization, $\lmax$, which is sensitive 
to the size of the aligned grains, typically lie in the range 0.5--0.6\mic\ 
for stars with $A_V < 3$, consistent with the mean value of 0.55\mic\ 
found in the diffuse ISM. Thus, the aligned grain population seems to 
behave normally in lines of sight lacking dense material, consistent 
with normal values of $R_V$ determined in the same extinction range. 
HD~283809 and (especially) HD~29647 have higher $\lmax$ values 
(see Table~1), consistent with the presence of larger grains in the 
clump component. 

We adopt the A2\,V star HD~283812 as the control for our study of 
interstellar polarization towards HD~29647 and HD~283809. There are 
several reasons why HD~283812 is a suitable standard. It is situated 
$\sim 45'$ SE of HD~29647 and HD~283809, in a region devoid of strong 
molecular emission. It has relatively low visual extinction 
($A_V\approx 1.9$), yet is quite highly polarized: the polarization 
efficiency, $\pmax/A_V \approx 3.3$~\%/mag, is close to the maximum value 
observed anywhere in the ISM, indicating a lack of depolarization effects 
associated with complex magnetic field structure or multiple cloud 
components (Gerakines et al\ 1995). Low/normal values of $R_V$ and $\lmax$ 
(Table~1) indicate that grain properties in the line of sight seem to be 
typical of the diffuse ISM. Most importantly, there is an absence of any 
detectable rotation with wavelength in the position angle of polarization 
towards HD~283812 (see Table~1 and Figure~1b): the standard deviation 
$\sigma_\theta$ in the mean is comparable with the observational uncertainty 
($\sim 1\deg$) in the individual measurements, and the slope 
$d\theta/d\lambda$ of a linear fit to the $\theta(\lambda)$ plot is not 
significantly different from zero. The mean position angle $<\theta>$ is 
consistent with the general trend of polarization vectors on the sky, which 
runs approximately NE--SW, as illustrated in Figure~2. More generally, this 
map strongly suggests the existence of a relatively uniform layer of 
polarizing dust permeating the region. The degree, position angle and 
wavelength-dependence of polarization towards HD~283812 are similar to 
those of other stars obscured by this layer (e.g., compare data for 
HD~283812 and Elias~19 in Whittet \etal\ 1992). Taking all of these factors 
into consideration, we conclude that HD~283812 is a reliable probe of grain 
properties in the extended layer; we assume HD~29647 and HD~283809 to have 
additional extinction arising in local concentrations where grain and 
magnetic field properties are different.

\section{ANALYSIS}
\subsection{Data and Empirical Fits}
The observational data used for this analysis consist of broadband linear 
polarimetry ($p$, $\theta$) in eight standard passbands between 0.35 and 
2.0\mic\ for each of the three program stars, taken from Whittet \etal\ 
(1992). The reader is referred to that paper for details of the 
individual measurements and discussion of data acquisition and reduction 
procedures. Plots of the degree, $p(\lambda)$, and position angle, 
$\theta(\lambda)$, of polarization against $\lambda^{-1}$ appear in 
Figure~1. Empirical fits to the data based on the Serkowski formula 
\begin{equation}
\frac{p(\lambda)}{\pmax}=\exp\left\{-K \ 
\ln^2\left(\frac{\lmax}{\lambda}\right)\right\}
\end{equation}
allow evaluation of the parameters $\pmax$ (the amplitude of peak 
polarization), $\lmax$ (the wavelength at which the peak occurs), and $K$ 
(sensitive to the width of the polarization curve). Fits performed by 
Whittet \etal\ (1992) are plotted in Figure~1a and values of the fit 
parameters are listed in Table~1. 

The polarization curves plotted in Figure~1a and represented by 
the Serkowski parameters in Table~1 refer to the observed $p(\lambda)$ 
{\it integrated over the entire pathlength to each star}. Our analysis 
below allows us to separate diffuse and dense components towards HD~29647 
and HD~283809 and to deduce independent Serkowski curves for the dense 
component.

\subsection{Description of Model}
We will test the theory that the lines of sight to HD~29647 and HD~283809 
contain distinct dense-cloud components which are responsible for anomalous 
behavior in the net observed polarization. It seems plausible that these 
dense regions are condensations embedded in the main TDC complex (although 
our analysis does not depend on this). A schematic representation of the 
adopted model is shown in Figure~3. For convenience, we refer to the 
proposed dense-cloud components in the lines of sight to HD~29647 and 
HD~283809 as ``Cloud~2a'' and ``Cloud~2b'', respectively. Although 
Clouds~2a and 2b are in all probability physically connected, the optical 
properties of material along these two lines of sight are not identical 
(\S~2). The extended cloud layer sampled by HD~283812 is referred to as 
``Cloud~1'', and is assumed to contribute equally to the extinction of all 
three stars. Thus, whereas the polarization towards HD~283812 is produced 
only by Cloud~1, the polarization towards HD~29647 is a superposition of 
the two polarized components (1 and 2a), and similarly HD~283809 (1 and 2b).

\subsection{Stokes Parameters}
The Stokes parameters, $Q$ and $U$, are related to the linear polarization,
$p$, by
\begin{equation}
Q/I=p\cos\,2\theta
\end{equation}
and 
\begin{equation}
U/I=p\sin\,2\theta
\end{equation}
where $p$ is the degree of linear polarization, $\theta$ is the position angle 
of polarization, and $I$ is the total intensity $I_{\rm max} + I_{\rm min}$
(Hall \& Serkowski 1963; Whittet 1992).

Our analysis relies on the fact that, for small
polarizations, the quantities $q = Q/I$ and $u = U/I$ are
additive over individual cloud components.
An approximate solution of the transfer equations
for partially polarized radiation (Martin 1974) shows that the
starlight transmitted by a series of $N$ uniform slabs
has Stokes parameters such that
\begin{equation}
q = \sum_{k=1}^N\,p_k\,\cos\,2\theta_k
\end{equation}
and
\begin{equation}
u = \sum_{K=1}^N\,p_k\,\sin\,2\theta_k,
\end{equation}
where $p_k$ and $\theta_k$ are, respectively, the
magnitude and position angle of the polarization that would
be observed if only the $k$th slab were present.
The preceding expressions are accurate to order $p^2$;
this is an excellent approximation here, where $p \sim 10^{-2}$.

Applying this result to our two component model, we have
\begin{equation}
q_{obs}=q_{1}+q_{2}
\end{equation}
and
\begin{equation}
u_{obs}=u_{1}+u_{2}
\end{equation}
where subscript 1 represents Cloud~1 and subscript 2 represent Cloud~2 
(a or b). Assuming $q_{1}$ and $u_{1}$ are given by observations of 
HD~283812 at a given wavelength, we calculate $q_{2}$ and $u_{2}$ 
from the observed polarization towards HD~29647 and HD~283809. This 
procedure is repeated for each available passband, allowing us to retrieve 
$p(\lambda)$ and $\theta(\lambda)$ for the dense cloud components in each 
line of sight. 

\section{RESULTS}

Results of the calculations described in \S~3.3 are given in Table~2.
The values listed are calculated $p(\lambda)$, $\theta(\lambda)$ 
for HD~29647 (Cloud~2a) and HD~283809 (Cloud~2b).
Uncertainties in the calculated quantities $p$ and $\theta$ were determined from the observational uncertainties. It is striking that the calculated $\theta(\lambda)$ 
for each cloud is almost independent of wavelength to within the calculated 
error bars (Figure~4b)\footnote{The large uncertainty in the calculated 
angle at 2.04\mic\ for the line of sight to HD~283809 is due to propagation 
of the uncertainty of the original measurement at this wavelength.}, consistent 
with our hypothesis that a simple two-component model adequately describes 
cloud and magnetic field structure along these lines of sight.

Taking a simple weighted average of the $\theta$ values in Table~2, we 
calculate the mean position angle of polarization for the cloud towards 
HD~29647 (cloud 2a) to be $\theta = 110 \deg \pm 1 \deg$, and similarly, 
for HD~283809 (cloud 2b), $\theta = 82 \deg \pm 2 \deg$. The dense cloud 
components for these stars are plotted as dashed lines in the visual 
polarization map in Figure~2. It is evident that the magnetic field 
direction in the dense material implied by our calculations is not consistent 
with the general trend for the extended region ($\theta \sim 30\deg$); 
towards HD~29647, it is nearly orthogonal. However, additional support for 
a substantially larger value for $\theta$ in the denser material is provided 
by near infrared (1.65~\micron) polarization measurements for highly 
obscured field stars in this region, reported by Moneti \etal\ (1984) 
(see their Figure~1). In particular, the object Elias~16 ($A_V \sim 20$~mag; 
Elias 1978) has $\theta \sim 78\deg$ at 1.65~\micron, similar to our 
calculated value for Cloud~2b. 

The calculated $p(\lambda)$ values in Table~2 have the expected form
for interstellar polarization and are well-represented by the Serkowski 
law (equ.~1). Fitting this formula to the data enables us to infer properties 
of the grain populations within the dense cloud material.
Results are shown in Figure~4a, and the fit parameters are listed in Table~3.
The data for HD~283812 are shown for comparison. We see that $\lmax$ values 
are longer for both Cloud~2a and Cloud~2b compared with Cloud~1 (Table~3), 
implying proportionately larger average grain radii, as expected for denser 
material where grain growth is most rapid (e.g., Jura 1980; Whittet 1992). 
That the largest value of $\lmax$ occurs towards HD~283809
is consistent with its higher extinction, implying a greater contribution 
from high density material. The lower peak polarization 
for Cloud~2b was not anticipated: this suggests that the polarizing
grains may not be as efficiently aligned in this region (again consistent 
with higher density; Goodman \etal\ 1995; Gerakines \etal\ 1995), but it
could also result from a less favorable magnetic field geometry within 
the dense cloud. The relatively high observed (net) polarization towards 
HD~283809 appears to arise predominantly in Cloud~1. Assuming that the 
dense component contributes extinction $[A_V]_{\rm dense} \sim 3.9$ towards 
HD~283809 (the difference between the total extinctions of HD~283809 and 
HD~283812), then the polarization efficiency in the dense material is 
$[\pmax/A_V]_{\rm dense} \sim 1.1$, a factor of 3 less than in the diffuse 
material sampled by HD~283812. 

Circular polarization is expected when the direction of grain alignment 
varies along the line of sight (Martin 1974). To estimate the magnitude $V/I$ 
of circular polarization predicted by our model, we integrated the radiative 
transfer equations for the Stokes parameters with respect to distance along 
each line of sight (Lee \& Draine 1985). We adopted a simple geometrical model 
consisting of three uniform slabs, where the foreground and background slabs 
represent the large-scale cloud and the intervening slab represents the 
embedded clump. We modeled the grains as perfectly aligned oblate spheroids 
composed of astronomical silicate (Draine \& Lee 1984), with axial ratios 
of $2$:$1$ (the shape inferred for aligned grains in nearby molecular clouds; 
Hildebrand \& Dragovan 1995). We assumed, somewhat arbitrarily, that the 
grains in the foreground and background slabs have radii $a=0.1$\mic; 
grains in the middle slab (the dense clump) are assumed to be larger by a 
factor $\eta$, where $\eta$ is the ratio of $\lmax$ inferred for the dense 
clump to that observed in the extended cloud toward HD~283812 (\S~4). 
In general, imperfect grain alignment and variations in the
magnetic field along the line of sight reduce the polarization by a 
factor $\Phi < 1$ relative to that observed for perfect grain alignment 
and a homogeneous magnetic field with the most favorable inclination.
For each line of sight, we estimated $\Phi$ by comparing the linear
polarization computed from our radiative transfer model (\S~4) to the linear
polarization observed in the line of sight.

Our calculations of the circular polarization predict detectable 
levels of circular polarization in the lines 
of sight toward HD~29647 and HD~283809. For example, in the $J$ (1.2\mic) 
passband, the calculations described in \S~3.4 indicate $V/I\sim 0.04$\% 
toward HD~29647 and $\sim 0.6$\% toward HD~283809. For comparison, Martin 
\& Angel (1976) observe $V/I$ in the range 0.02--0.08\% in the near infrared 
towards several reddened stars. A search for circular polarization toward 
HD~29647 and HD~238809 would provide a test of our model.

\section{CONCLUSION}

The observed rotation of the polarization position angle with wavelength
towards HD~29647 and HD~283809 is consistent with a two-component model
for polarizing material in these lines of sight, in which the net 
polarization is produced by distinct cloud regions with differing average 
magnetic field direction and grain size. The most plausible model for the 
region is a sheet of continuous material extending over much of the Taurus 
dark cloud, with an embedded dense core (or cores) in the lines of sight to 
HD~29647 and HD~283809. Analysis based on the Stokes' parameter allows us 
to estimate the polarization properties of the dense components. The 
wavelength dependence of linear polarization is shown to be well-represented 
by the Serkowski empirical formula with $\lmax \approx 0.61$\mic\ (HD~29647) 
and 0.73\mic\ (HD~283809), compared with 0.55\mic\ for the continuous sheet. 
The position angle of polarization for the dense material towards HD~29647 
appears to be almost perpendicular to that of the cloud as a whole. 
Detectable levels of circular polarization are predicted toward HD~29647 
and HD~283809, and a future search for circular polarization in these lines of 
sight would provide a useful test of our model.

\acknowledgements
We would like to thank Lida He for the use of his Serkowski fitting program. 
We also thank Don Mizuno and Sean Carey for assistance with the data 
analysis package REDUCE used in the creation of the polarization map.
This research is funded by NASA grants NAGW-3001 and NAGW-3144.

\onecolumn
\begin{figure}[p] 
\figcaption{Observed spectral dependence of polarization (degree and 
position angle) towards HD~29647 (squares), HD~283809 (circles) and 
HD~283812 (triangles). The curves fit to the $p(\lambda)$ data 
(Figure~1a) are based on the Serkowski formula (equation~1) with 
$\pmax$, $\lmax$ and $K$ treated as free parameters. The lines in 
Figure~1b are linear least squares fits to $\theta(\lambda)$.} 
\psfig{figure=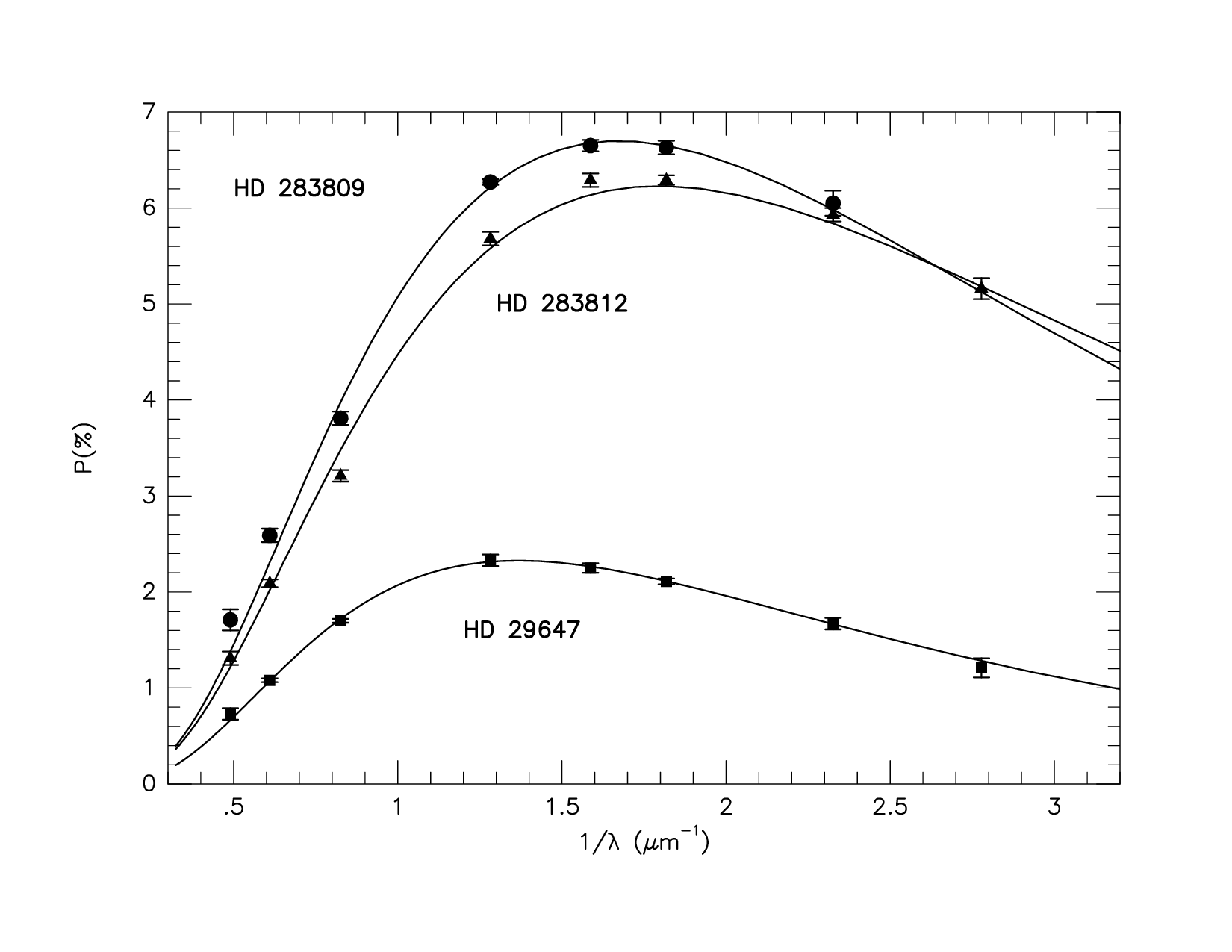,height=8in,rheight=8in,angle=90}
\end{figure}

\begin{figure}
\psfig{figure=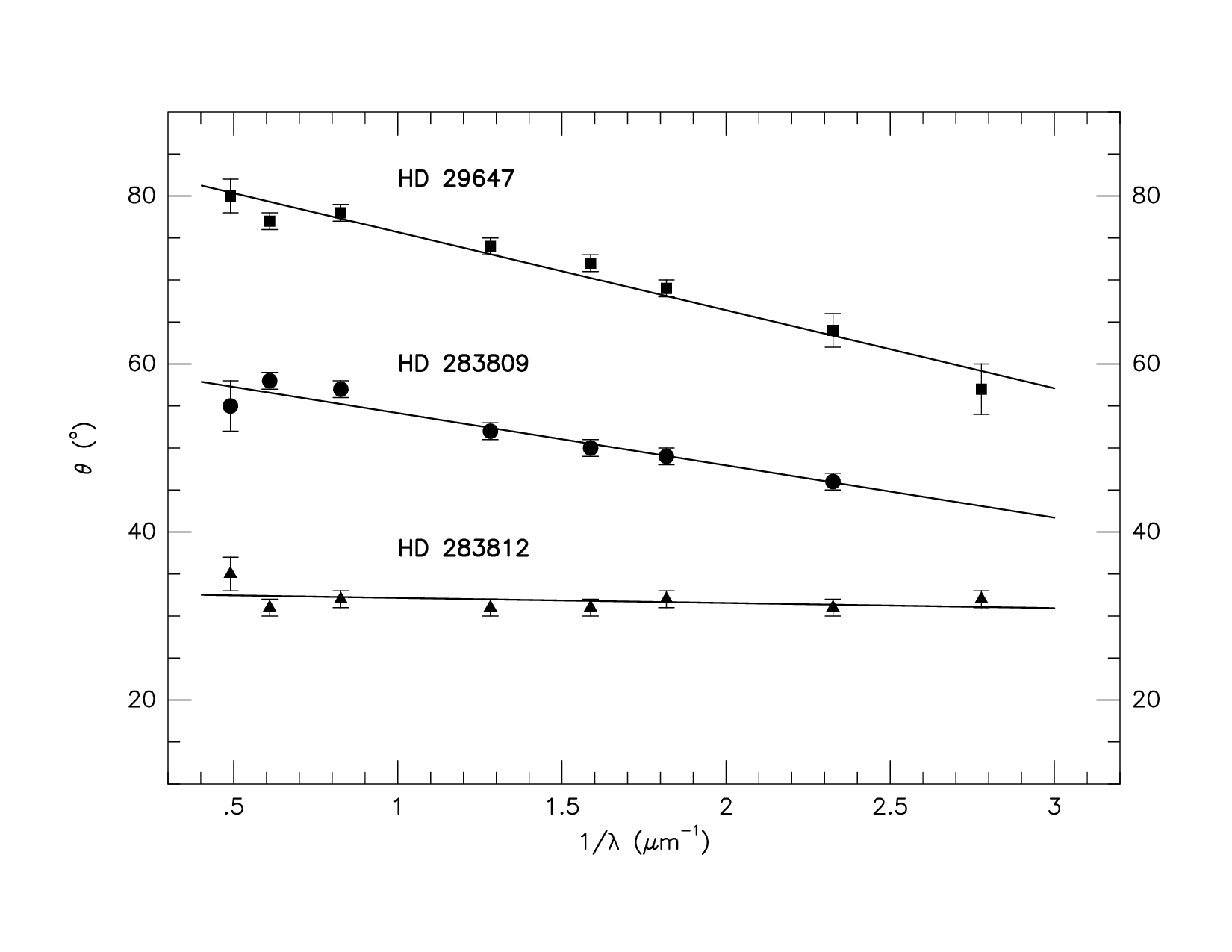,height=8in,rheight=8in,angle=90}
\end{figure}

\begin{figure}[p] 
\figcaption{Linear polarization map of the Taurus dark cloud region 
surrounding HD~29647. Field stars from Moneti \etal\ (1984) are denoted 
by $+$ signs and their visual polarization vectors represented by solid 
lines. The stars HD~29647, HD~283809 and HD~283812 (in sequence from North 
to South) are plotted as filled circles using data from Whittet \etal\ (1992).  
Observations of the line of sight, and calculated values for the modeled 
clumps to HD~29647 and HD~283812 (see text), are represented by
solid and dashed lines, respectively.}
\centerline{\psfig{figure=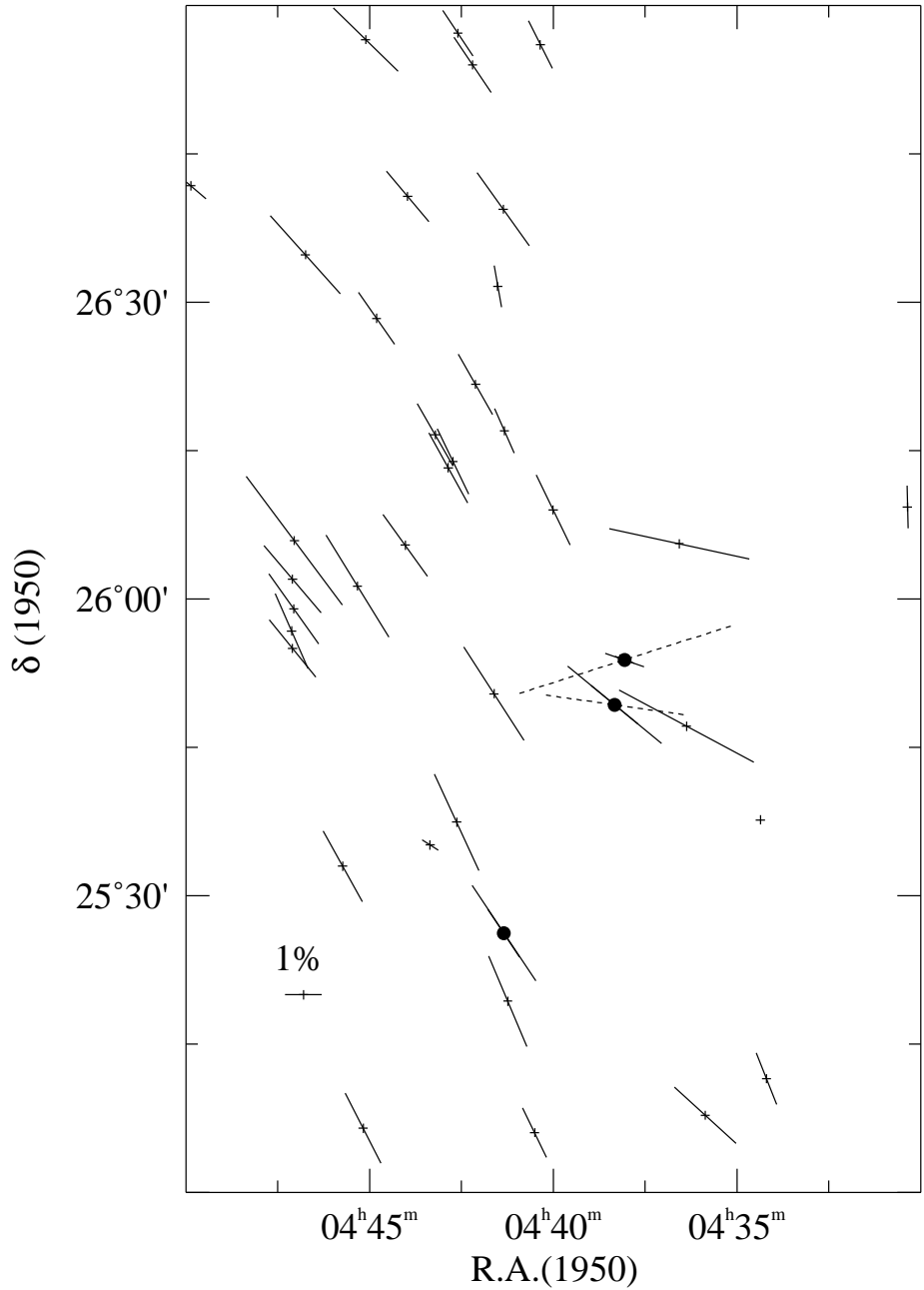,height=7in,rheight=7.5in}}
\end{figure}

\begin{figure}[p] 
\figcaption{Schematic representation of our model for the TMC--1 region, 
in which a continuous sheet of material (sampled by HD~283812, Cloud~1) 
contains denser embedded concentrations. Dense material towards HD~29647 
and HD~283809 is labelled Cloud~2a and Cloud~2b, respectively.}
\psfig{figure=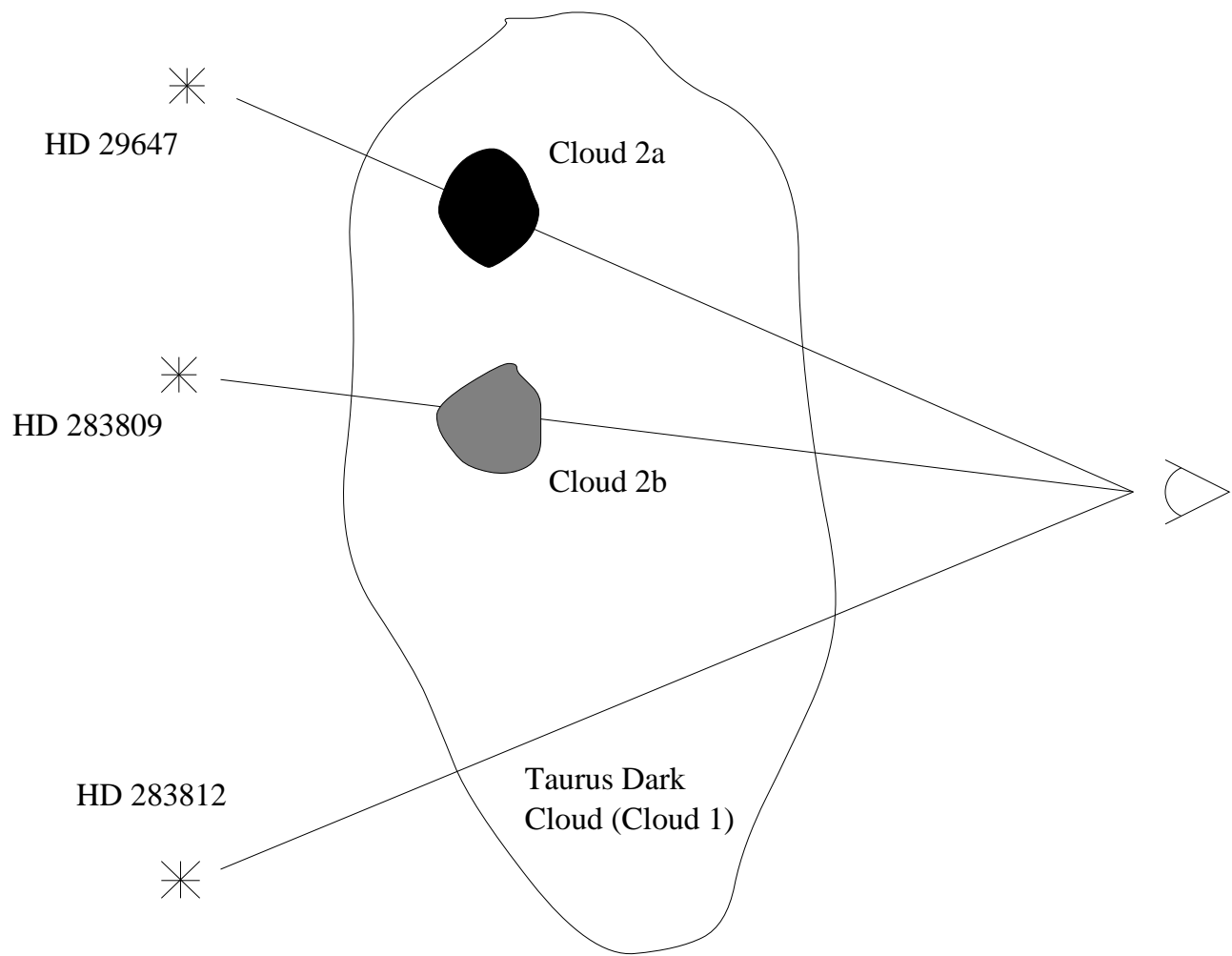,height=6in,rheight=7in,angle=-90}
\end{figure}

\begin{figure}[p] 
\figcaption{Similar to Figure~1, but showing calculated polarization 
data (degree and position angle) towards HD~29647 (Cloud~2a) and 
HD~283809 (Cloud~2b). Data for HD~283812 (Cloud~1) are shown for 
comparison. The parameters of the Serkowski curves fit to $p(\lambda)$ 
are given in Table~3.}
\psfig{figure=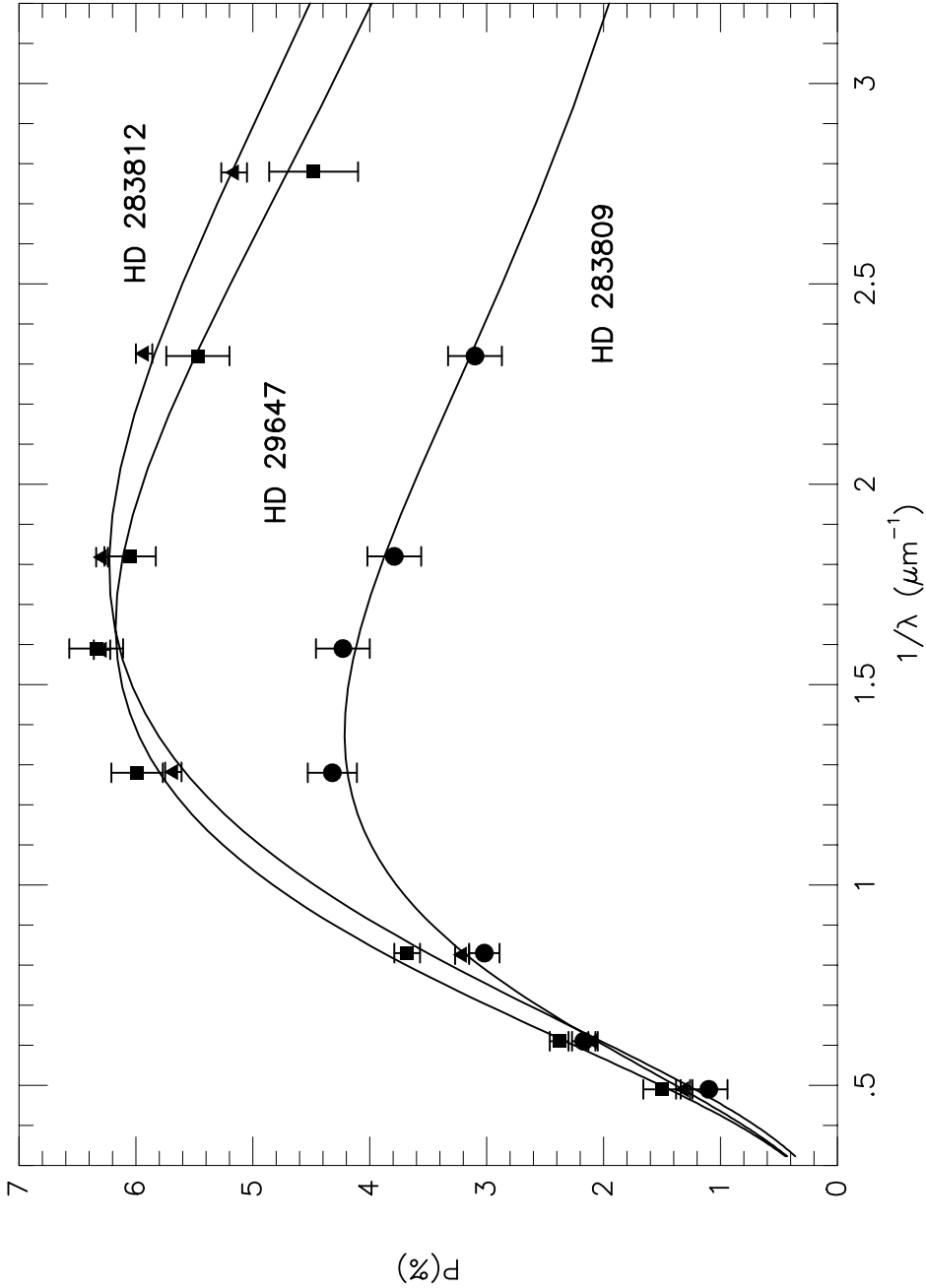,height=8in,rheight=10in,angle=90}
\end{figure}
\begin{figure}
\psfig{figure=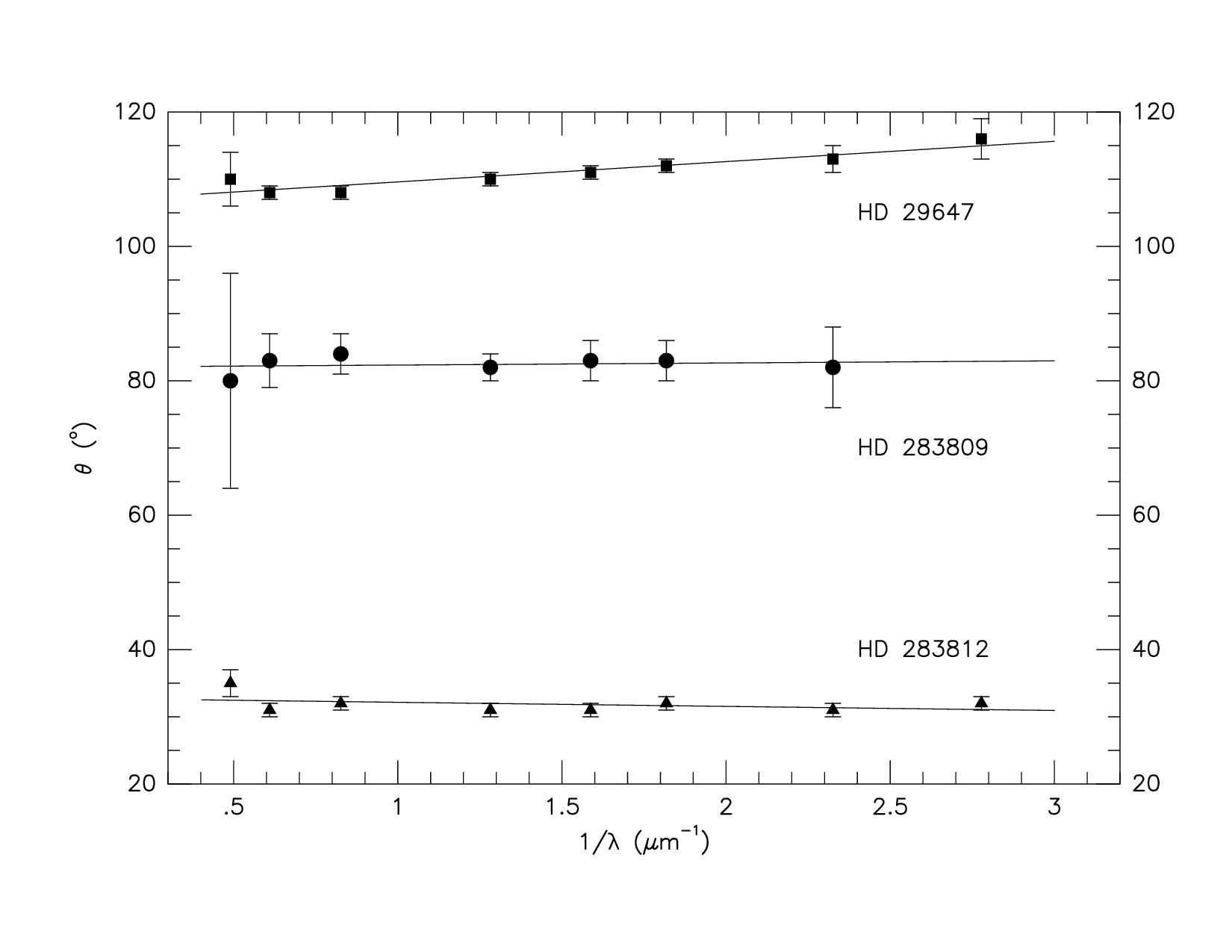,height=8in,rheight=10in,angle=90}
\end{figure}

\begin{table} 
\begin{center}
\caption{SPECTRAL TYPE, EXTINCTION AND POLARIZATION DATA FOR THE PROGRAM STARS}
\begin{tabular}{cccc}
\\
\hline\hline
&HD~29647&HD~283809&HD~283812 \\ 
\hline
\\
Spectral Type &B6-7 IV\tablenotemark{a}&B3 V\tablenotemark{b}
&A2 V\tablenotemark{c} \\
$E_{B-V}$ &1.00\tablenotemark{a}&1.61\tablenotemark{b}
&0.66\tablenotemark{c} \\
$R_{V}$ \tablenotemark{d}&3.62&3.58&2.87 \\
$p_{\rm max}$ (\%)\tablenotemark{e} &2.30&6.70&6.30 \\
$\lambda_{\rm max}$ ($\micron$)\tablenotemark{e} &0.73&0.59&0.55 \\
$K$ \tablenotemark{e} &1.15&0.97&0.96 \\
$<\theta>$ ($\deg$)\tablenotemark{e} &73.5&52.0&31.6\\
$\sigma_{\theta}(\deg)$\tablenotemark{e} &7.8&4.4&1.4 \\ 
$d\theta/d\lambda$ ($\deg/\micron$)\tablenotemark{e} & 13.9 &7.3 &2.4 \\
\\
\hline
\end{tabular}
\tablenotetext{a}{Vrba \& Rydgren 1985; Crutcher 1985.}
\tablenotetext{b}{Strai\v{z}ys \etal\ 1985.}
\tablenotetext{c}{Strai\v{z}ys \& Meistas 1980.}
\tablenotetext{d}{Gerakines \etal\ 1995.}
\tablenotetext{e}{Whittet \etal\ 1992.}
\end{center}
\end{table}

\begin{table} 
\begin{center}
\caption{CALCULATED LINEAR POLARIZATIONS FOR CLOUD COMPONENTS 2a AND 2b}
\begin{tabular}{cccccccccc}
\\
\hline
\hline
&\multicolumn{4}{c}{HD~29647  (Cloud~2a)} &
&\multicolumn{4}{c}{HD~283809 (Cloud~2b)}\\
~~~~$\lambda$ ($\micron$)~~~~&
$p$ (\%)&$\delta p$ (\%)&$\theta$ $(\deg)$&$\delta \theta$ $(\deg)$ &&
$p$ (\%)&$\delta p$ (\%)&$\theta$ $(\deg)$&$\delta \theta$ $(\deg)$ \\ 
\hline
\\
0.36  &  4.48  &  0.38  &  116  &  3 \\
0.43  &  5.47  &  0.27  &  113  &  2 &&  3.10  &  0.23  &  82  &  6\\
0.55  &  6.05  &  0.22  &  112  &  1 &&  3.79  &  0.23  &  83  &  3\\
0.63  &  6.34  &  0.23  &  111  &  1 &&  4.23  &  0.23  &  83  &  3\\
0.78  &  5.99  &  0.22  &  110  &  1 &&  4.32  &  0.21  &  82  &  2\\
1.21  &  3.68  &  0.11  &  108  &  1 &&  3.02  &  0.13  &  84  &  3\\
1.64  &  2.38  &  0.08  &  108  &  1 &&  2.17  &  0.10  &  83  &  4\\
2.04  &  1.50  &  0.16  &  110  &  4 &&  1.10  &  0.19  &  80  &  16\\ 
\\
\hline
\end{tabular}
\end{center}
\end{table}

\begin{table} 
\begin{center}
\caption{SERKOWSKI FIT PARAMETERS FOR HD~283812 (CLOUD~1), HD~29647 (CLOUD~2a) AND HD~283809 (CLOUD~2b)}
\begin{tabular}{cccc} 
\\
\hline
\hline
Parameter&HD~283812 (Cloud~1)&HD~29647 (Cloud~2a)&HD~283809 (Cloud~2b) \\ 
\hline
\\
$\pmax$&$6.30\pm0.07$&$6.17\pm0.24$&$4.22\pm0.25$\\
$\lmax$&$0.55\pm0.01$&$0.61\pm0.04$&$0.73\pm0.05$\\
$K$    &$0.96\pm0.05$&$0.99\pm0.15$&$1.08\pm0.24$\\
\\
\hline
\end{tabular}
\end{center}
\end{table}
\end{document}